\documentclass{article}
\usepackage{authblk}
\usepackage[utf8]{inputenc}
\usepackage{geometry}
\usepackage{graphicx}
\usepackage{cite}
\usepackage{amsmath}
\usepackage{subfigure}
\usepackage{color}

\usepackage{latexsym}

\usepackage{xspace}

\newcommand{\um}{$\mu$m\xspace}
\newcommand{\ums}{$\mu$m$^2$\xspace}
\newcommand{\Celsius}{$^\circ$C\xspace}
\newcommand{\sinx}{Si$_3$N$_4$\xspace}
\newcommand{\siox}{SiO$_2$\xspace}

\begin{document}
\title{Thermal tuners on a Silicon Nitride platform}

\author[1]{Daniel P\'erez}
\author[2]{Juan Fern\'andez}
\author[1]{Roc\'io Ba\~nos}
\author[2]{Jos\'e~David~Dom\'enech}
\author[3]{Ana M. Sánchez}
\author[3]{Josep M. Cirera}
\author[3]{Roser Mas}
\author[3]{Javier Sánchez}
\author[3]{Sara Durán}
\author[3]{Emilio Pardo}
\author[3]{Carlos Dom\'inguez}
\author[1]{Daniel Pastor}
\author[1]{Jos\'e Capmany}
\author[1,2]{Pascual~Mu\~noz\footnote{Corresponding author: P. Mu\~noz, e-mail: pascual@ieee.org.}}
\affil[1]{Photonic IC-group at the Photonics Research Labs, Universitat Polit\`ecnica de Val\`encia, C/ Camino de Vera s/n, Valencia 46022, Spain. e-mail: pascual@ieee.org.}
\affil[2]{VLC Photonics S.L., C/ Camino de Vera s/n, Valencia 46022, Spain. e-mail: david.domenech@vlcphotonics.com}
\affil[3]{Instituto de Microelectrónica de Barcelona (IMB-CNM, CSIC) C/ del Til·lers. Campus Universitat Autònoma de Barcelona (UAB) 08193 Cerdanyola del Vallès e-mail: carlos.dominguez@imb-cnm.csic.es}


\maketitle

\begin{abstract}
In this paper, the design trade-offs for the implementation of small footprint thermal tuners on silicon nitride are presented, and explored through measurements and supporting simulations of a photonic chip based on Mach-Zehnder Interferometers. Firstly, the electrical properties of the tuners are assessed, showing a compromise between compactness and deterioration. Secondly, the different variables involved in the thermal efficiency, switching power and heater dimensions, are analysed. Finally, with focus on exploring the limits of this compact tuners with regards to on chip component density, the thermal-cross talk is also investigated. Tuners with footprint of 270x5~\ums and switching power of 350~mW are reported, with thermal-cross talk, in terms of induced phase change in adjacent devices of less than one order of magnitude at distances over 20~\um. Paths for the improvement of thermal efficiency, power consumption and resilience of the devices are also outlined. 
\end{abstract}


\section{\label{sec:intro}Introduction}
Photonic integrated circuits (PICs) have gained enormous momentum in the recent years as fundamental blocks for multiple applications domains, from telecom/datacom, through bio/life sciences, avionics/aeronautics, safety, security to civil engineering and construction \cite{proc-ieee-willner}. PICs are becoming ubiquitous due to the push and rapid evolution from the mainstream integration technologies, namely Silicon-on-Insulator (SOI) \cite{tech-soi-lim,tech-soi-lo} and Indium Phosphide (InP) \cite{tech-inp-iop,tech-inp-smit,tech-inp-kish}. 

Alongside, Silicon Nitride based integration platforms are subject of attention due to the wide wavelength range over which the material is transparent (400-2400~nm) and inherently low-loss. However, dielectric based photonic technologies started two decades ago, with the development of components in the visible wavelength range for optical sensors \cite{tech-sinx-seminal-1,tech-sinx-seminal-2}. This waveguide technology is based on a combination of stoichiometric silicon nitride (\sinx) as waveguide layers, filled by and encapsulated with silica (\siox) as cladding layers grown on a silicon wafer \cite{tech-sinx-leinse,tech-sinx-witzens,tech-sinx-domenech}. \siox and \sinx layers are fabricated with CMOS-compatible industrial standard chemical vapor deposition (both low pressure, LPCVD, and plasma enhanced, PECVD) techniques, that enable cost-effective volume production.

Aiming at the reconfiguration of PICs, various physical mechanisms exist and are present in the different technology platforms \cite{tech-soi-lim,tech-soi-lo,tech-inp-iop,tech-inp-smit,tech-inp-kish,tech-sinx-leinse,tech-sinx-witzens,tech-sinx-domenech}. Electro-absorption and electro-refraction are faster and use less energy, however the thermo-optic effect over the refractive index is larger, and induces less losses \cite{tuning-methods}. Despite their main drawbacks, power consumption and thermal cross-talk, and the proposed improvement alternatives \cite{tech-sinx-piezo,tech-inp-trenches} resorting to additional process steps, regular thermal tuners are simple and commonplace. However, most of the approaches in the literature, for all technologies, propose the use of long and wide tuners for linear phase-shift vs. driving power operation: in \cite{tech-sinx-mwp}, a heater length of 2~mm is given, in \cite{heater-data-sinx-dmux}, a 40~mm length tuner is presented and in \cite{heater-data-hybrid}, lengths above 600~\um are employed, to cite a few. In terms of the length, long heaters are contrary to the spirit of PICs, where footprint ultimately determines the cost of the photonic circuit. Furthermore, a common given figure given in the literature is the switching power to obtain a phase shift of $\pi$, namely $P_\pi$. For instance, in \cite{heater-data-sinx-dmux} with a technology similar to one being in this paper, a switching power of 350~mW, corresponding to a temperature increase of 40~\Celsius is reported. However, little information on the internal construction of a very basic element as a regular thermal tuner is given. Since, as it will be shown in this paper, not only $P_\pi$ counts, but also the heater and optical waveguide cross-sections, it is not straightforward to compare just with switching powers. Moreover, for increased component density on chip, the thermal cross-talk is an important issue, not otherwise present, or reported, previously in the literature with long heaters. Therefore, for ultra-compact circuits, a clear understanding of electric and thermo-optic properties of thermal tuners is required. 

In this paper, the design, fabrication in \sinx/\siox technology, experimental demonstration, and complementary supporting simulations for small footprint thermal tuners are provided. The paper is structured as follows. In section~\ref{sec:tech}, the technology platform and fabrication processes are briefly described. The electrical properties of thermal tuners in terms of their dimensions and driving power are reported in Section~\ref{sec:elec}. Next, the switching power for several structures is presented in Section~\ref{sec:ppi}, where the relations with the thermal efficiency and electrical properties from the previous section are established. In Section~\ref{sec:xt}, the thermal cross-talk for different tuners is reported, both for nearby and further apart optical structures. At the sight of all the results, paths for improvements are outlined in Section~\ref{sec:mejoras}. Finally, the conclusions are derived in Section~\ref{sec:conclusion}.

\section{\label{sec:tech}Technology platform}
\begin{figure}
  {\par\centering
   \resizebox*{0.48\textwidth}{!}{\includegraphics*{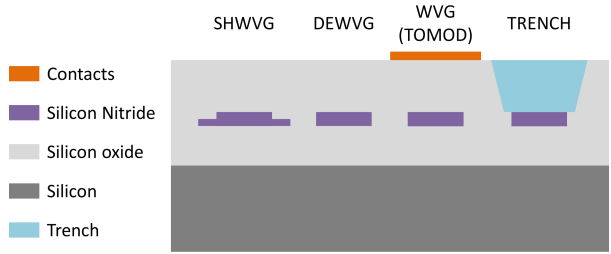}}
   \caption{Silicon nitride technology platform waveguide cross-sections}
  \label{fig:cs}
  }
\end{figure}

The designed devices were fabricated in a Silicon Nitride photonics platform compatible with a standard CMOS pilot line. Several regular and custom waveguide cross-sectional geometries are available in these platforms \cite{tech-sinx-leinse,tech-sinx-domenech}. The waveguide basic structures, shown in Fig.~\ref{fig:cs}, are fabricated on 100~mm silicon wafer diameter, and the layer stack is composed of a silicon oxide buffer layer (2.0~\um~thick, n=1.464) growed by thermal oxidation of the silicon substrate, following a LPCVD silicon nitride layer with a thickness of 300~nm (n= 2.01) acts as the core layer and finally a 1.5~\um~thick silicon oxide (n=1.45) cladding layer is deposited by PECVD \cite{tech-sinx-domenech}. Two different waveguide structures are defined by photo-lithography with a Stepper i-line (minimum feature 500~nm) followed by a reactive ion etching (RIE) of the silicon nitride film. The 300~nm silicon nitride layer may be etched completely to form a strip waveguide structure (DEWVG), or the etching is done partially obtaining a rib waveguide structure (SHWVG). Two additional processes are used for defining the heater and the air wells. A metal heater (TOMOD) is obtained by sequential evaporation of 30~nm Chromium and 100~nm Gold, and defined by a lift-off process. The air wells (TRENCH) are open into the cladding layer up to the silicon nitride core by photo-lithography followed by a RIE step. 

\begin{figure}
  {\par\centering
    \resizebox*{0.48\textwidth}{!}{\includegraphics*{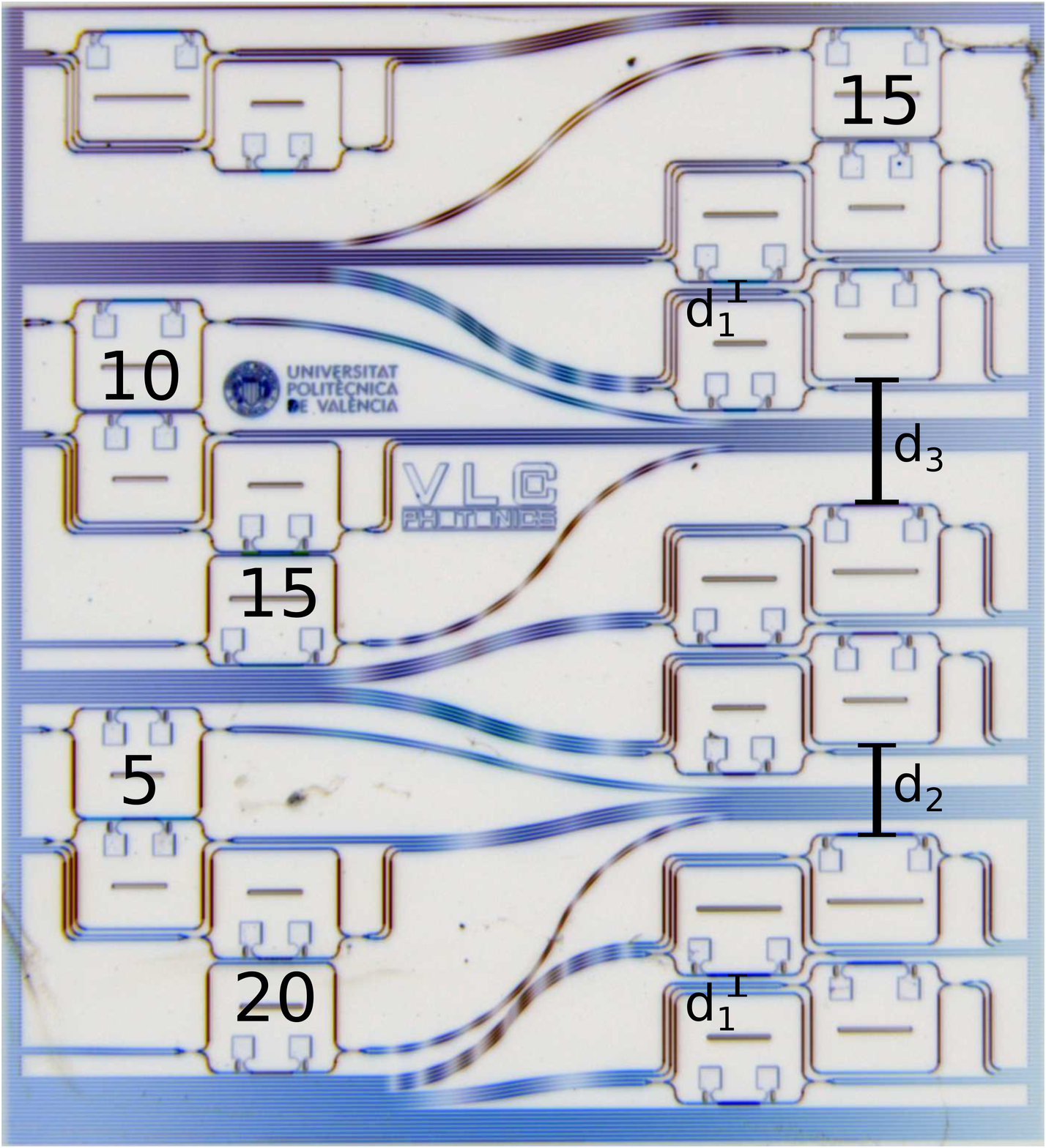}}
   \caption{Annotated micro-graph (inverted color) of MPW cell with MZI test structures. The numbers indicate the separation (\um) between close MZIs (die size 5.5x5.5~mm$^2$). The distances for MZIs further apart are $d_1=86$~\um, $d_2=426$~\um~and $d_3=586$~\um.}
  \label{fig:chip}
  }
\end{figure}

\begin{figure}
  {\par\centering
   \resizebox*{0.48\textwidth}{!}{\includegraphics*{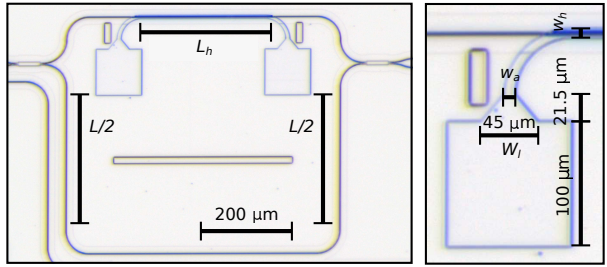}}
   \caption{Annotated micro-graph (inverted color) of asymmetric Mach-Zehnder Interferometer with thermo-optic tuner (left) and enlarged view of the pad and access metal track (right).}
  \label{fig:mzi}
  }
\end{figure}

To asses on the electrical and optical properties of the thermal tuners, a test die with Mach-Zehnder Interferometer (MZI) test structures was fabricated, and a micro-graph picture is shown in Fig.~\ref{fig:chip}. The MZIs are built by interconnecting two Multimode Interference (MMI) couplers, with two waveguides of different lengths. The thermal tuner is laid out over the shortest MZI arm, Fig.~\ref{fig:mzi}. The tuners are composed of three metal sections, named pads with tapers, curved access and heater (right panel at Fig.~\ref{fig:mzi}). The pads are 100x100~\ums. The taper section from pads to the access section is linear, starting at a width of $W_l=45$~\um~down to the access section width $w_a=w_h$+2~\um, where $w_h$ is the heater section width. This linear taper has a length of 21.5~\um. The access is a curved track, with radius $r_a=50$~\um, whose width is tapered from $w_a$ to $w_h$. The chip is populated with MZIs having heaters of widths 5, 6, 7 and 8~\um, and lengths of 120, 130, 170, 220, 270 and 320~\um. 
\section{\label{sec:elec}Electrical properties}

\begin{figure}
  {\par\centering
   \resizebox*{0.48\textwidth}{!}{\includegraphics*{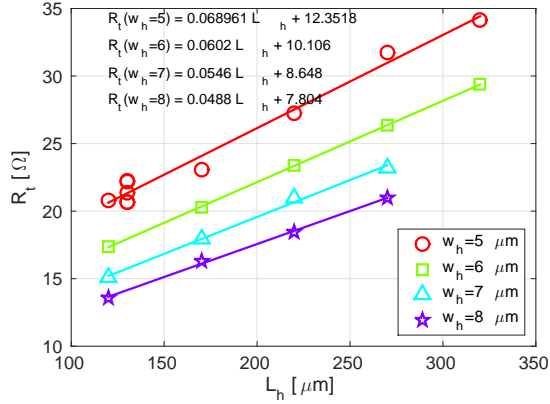}}
   \caption{Resistance for different heater widths and lengths.}
  \label{fig:A_RLw}
  }
\end{figure}

The equivalent electric circuit for the thermal tuner is composed of five resistors in series, corresponding to the two pads, the two tapers, the two accesses, and the heater sections. The resistance in $\Omega$ is given by the well known law:
\begin{equation}
\label{eq:R}
R = \rho \frac{L}{S} = \rho \frac{L}{w \cdot t} = R_S \frac{L}{w}
\end{equation}
where $\rho$ is the resistivity in $\Omega \cdot$m, $L, w, t$ are the length, width and thickness in m, and $R_s = \rho / t$ is the square sheet resistance in $\Omega / sq$. In the platform, $t=130$~nm total, composed of the two aforementioned layers of Cr and Au of 30~nm and 100~nm respectively. 
The resistance for the complete thermal tuner $R_t$ is then given by:
\begin{equation}
\label{eq:Rt}
R_t = R_h + 2 R_p + 2 R_l + 2 R_a 
\end{equation}
with 'h', 'p', 'l' and 'a' meaning heater, pad, linear taper and curved linearly tapered access respectively. In terms of the sheet resistance, the previous equation can be written as:
\begin{equation}
\label{eq:Rt2}
R_t = R_S \left( \frac{L_h}{w_h} + 2 + 2 F_l + 2 F_a \right) 
\end{equation}
where $F_l$ and $F_a$ correspond to the linear taper and curved linearly tapered access, and are given in the Appendix~\ref{app:res}. 
Fig.~\ref{fig:A_RLw} shows the resistance measurements for tuners with different heater lengths and widths (symbols), alongside a linear regression. The measurements were performed with a Keithley~2401 current source, operating at a constant DC current, and recording the different voltage readings. From eq. (\ref{eq:Rt}) we obtain an average resistivity of $\bar{\rho} = 4.77\cdot10^{-8}$~$\Omega \cdot$m, with standard deviation of $\sigma_\rho = 1.7\cdot10^{-9}$~$\Omega \cdot$m. Accordingly, the average square sheet resistance $\bar{R_S} = 0.368$~$\Omega/sq$. 
\begin{figure}
  {\par\centering
   \resizebox*{0.48\textwidth}{!}{\includegraphics*{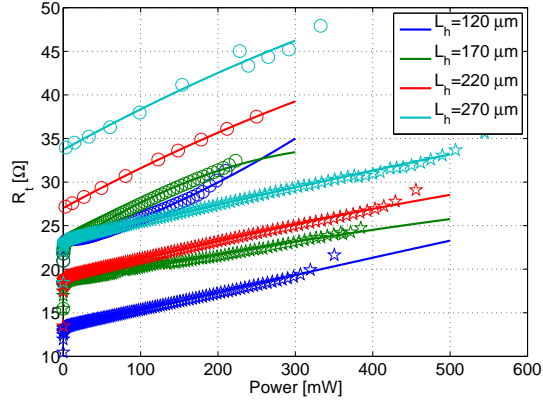}}
   \caption{Thermal tuner resistance increase with driving power, for widths $w_h=5$~\um~(circles) and $w_h=8$~\um~(stars), and different lengths}
  \label{fig:B_rho}
  }
\end{figure}
Resistivity is reported to grow with temperature \cite{heater-metales} up to a point for which the heater structure is destroyed. Fig.~\ref{fig:B_rho} shows the thermal tuner resistance increase with driving power, for the different heater lengths on chip, and for widths $w_h=5$~\um~and $w_h=8$~\um. This is connected to footprint and switching power in the next section.
\begin{figure}
  {\par\centering
    \subfigure[$w_h=8$~\um, $\Delta T_c\simeq 54$~\Celsius.]{\resizebox*{0.5\textwidth}{!}{\includegraphics*{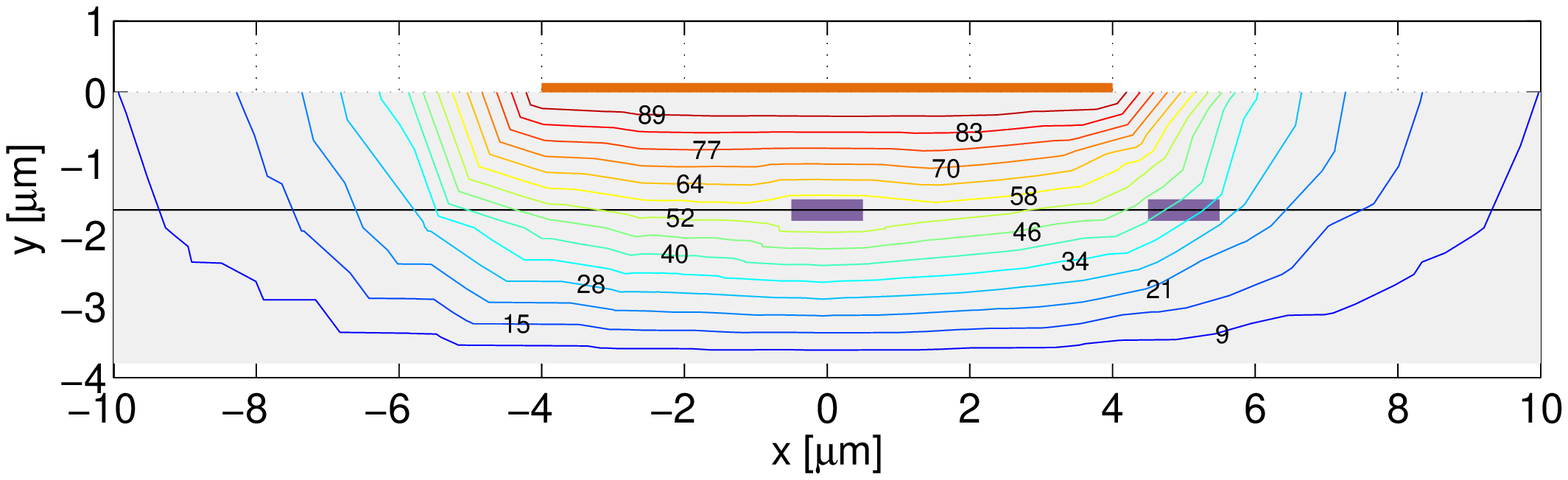}}}
    \subfigure[$w_h=7$~\um, $\Delta T_c\simeq 58$~\Celsius.]{\resizebox*{0.5\textwidth}{!}{\includegraphics*{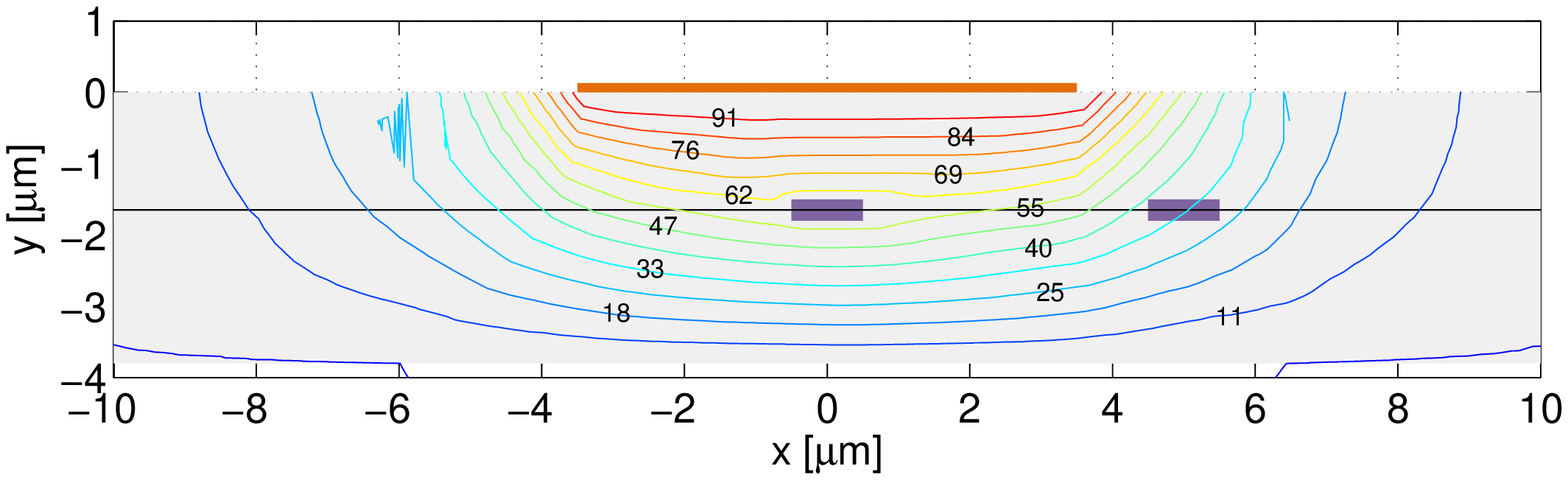}}}
    \subfigure[$w_h=6$~\um, $\Delta T_c\simeq 60$~\Celsius.]{\resizebox*{0.5\textwidth}{!}{\includegraphics*{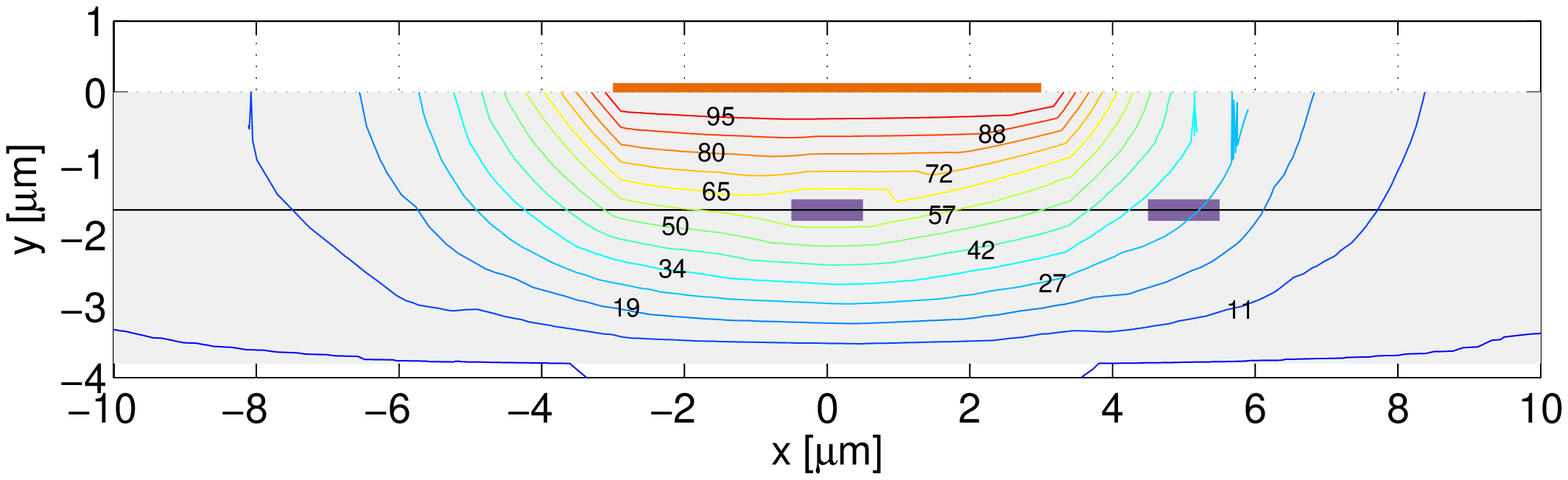}}}
    \subfigure[$w_h=5$~\um, $\Delta T_c\simeq 63$~\Celsius.]{\resizebox*{0.5\textwidth}{!}{\includegraphics*{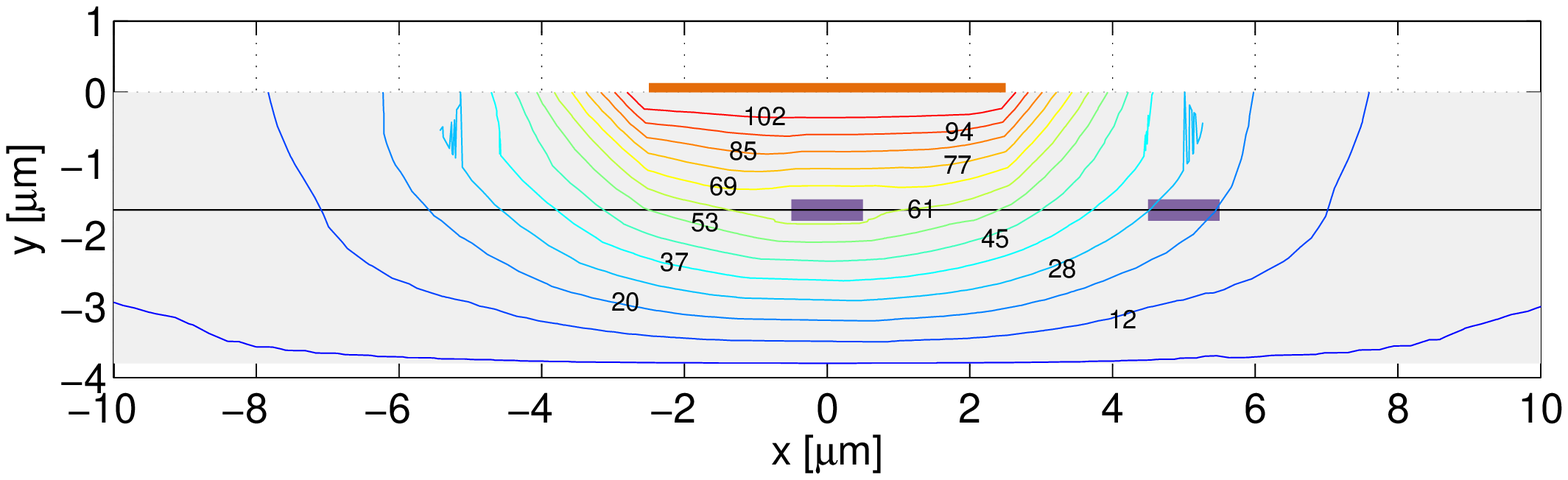}}}
   \caption{Waveguide cross-section with heater on top, and adjacent waveguide at a distance of 5~\um, with temperature gradient distribution overlaid. A dotted line is drawn from left to right crossing the core of the waveguides at half their height. Four different heater widths are shown, for the same heater length ($L_h=270$~\um). The contours are simulated for the same heater power consumption (same amount of heat generated), showing the temperature gradient is larger for narrow heaters.}
  \label{fig:E_contours}
  }
\end{figure}

\section{\label{sec:ppi}Switching power}
For a current flowing through the thermal tuner, the heater temperature is proportional to the current and the resistance. The heater is on top of an optical waveguide, and the heat will flow down towards the waveguide core. This is shown in Fig.~\ref{fig:E_contours}. The heat will alter the refractive indices of both the waveguide core and cladding, resulting into a change in the waveguide mode effective index. Therefore, a change in the optical phase of the propagated light will be imposed, which can be expressed as:
\begin{equation}
\label{eq:phi_pi}
\frac{\Delta \phi}{2\pi} = \frac{1}{\lambda}\Delta\bar{n}L_h^e = \frac{1}{\lambda} \frac{\partial \bar{n}}{\partial T_c} \Delta T_c\left(P,L_h,w_h\right) L_h^e
\end{equation}
where $\partial\bar{n}/\partial T_c$ is the rate of change of the effective index $\bar{n}$ vs the waveguide core temperature $T_c$, $\Delta T_c(P,L_h,w_h)$ is the core temperature change induced by a heater driven with power $P$, and $L_h^e$ is the heater effective length where the temperate change takes place. Using two different software packages (PhoeniX OptoDesigner and COMSOL Multi Physics) with the material properties (Appendix~\ref{app:mat}) and a waveguide cross-section for a strip waveguide (DEWVG),  $\partial \bar{n}/\partial T_c\simeq 3.05 \cdot 10^{-5}$~K$^{-1}$ was obtained. Moreover, the heat transfer efficiency from the heater metal to the core of the waveguide (for this particular cross-section) $\Delta T_c / \Delta T_h$ was computed to be approximately of 58\%. The data from the simulations was extracted in thermal steady state, and heat conduction was considered, since convection is negligible \cite{thermo-book}. 
 
The heater dimensions influence on the temperature gradient that is created in the cross-section. For the same amount of heat, that is for the same amount of electric consumed power at the heater, $P$, different temperature gradients result from heaters of different widths and lengths. This is stated explicitly as  $\Delta T_c(P,L_h,w_h)$ in eq. (\ref{eq:phi_pi}). In Fig.~\ref{fig:E_contours} four heaters with the same length, $L_h=270$~\um, and widths of 5, 6, 7 and 8~\um, are simulated for the same consumed power, with COMSOL~MP software package.  The silicon nitride waveguide, 1.0x0.3~\ums~sits on top of 2~\um~of buried oxide, which in turn was grown on top of a silicon wafer of thickness 500~\um. In the simulations, the temperature at the bottom part of the silicon is fixed to 25~\Celsius. The voltage is set in the simulator for the wider heater ($w_h=8$~\um) to $2.5$~V, and owing to Fig.~\ref{fig:A_RLw}, the resistance is $\simeq 21$~$\Omega$ ($P=297.62$~mW). For the narrow heaters, $V$ is set to provide the same power by using the resistance values from Fig.~\ref{fig:E_contours}. The resulting temperature gradient in the core of the silicon nitride waveguide is approximately 11~\Celsius~ more for the heater with narrower width. Hence, the same amount of heat is concentrated by narrow heaters, creating larger temperature gradients.

\begin{figure}
  {\par\centering
   \resizebox*{0.48\textwidth}{!}{\includegraphics*{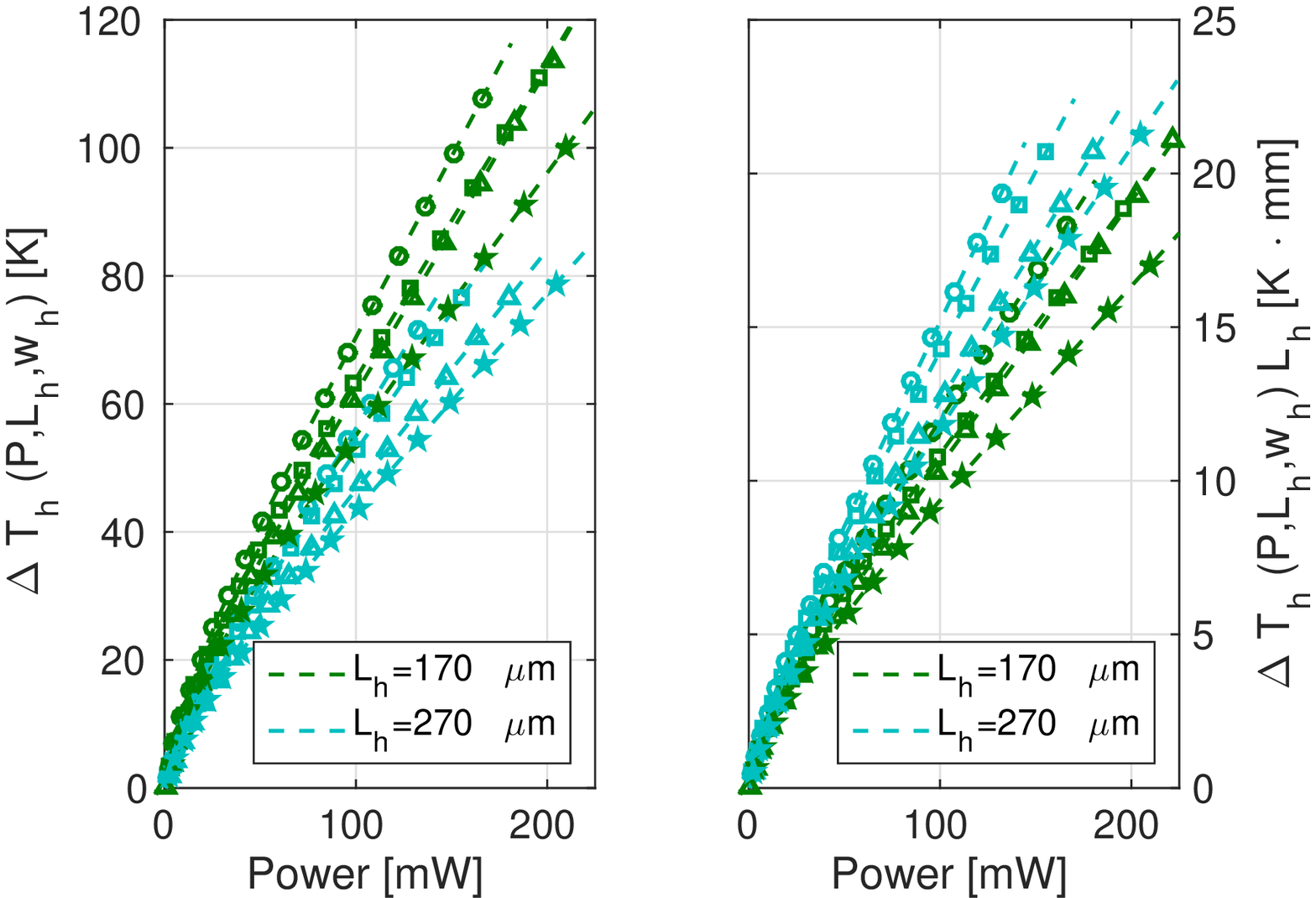}}
   \caption{ $\Delta T_h$ (left) and $\Delta T_h \cdot L_h$ (right) for widths $w_h=5$~\um~(circles), $w_h=6$~\um~(squares), $w_h=7$~\um~(triangles) and $w_h=8$~\um~(stars), for heaters lengths 170 and 270~\um.}
  \label{fig:ATh}
  }
\end{figure}

To infer more precisely on the variation of $\Delta T_c$ vs all the involved parameters, $P$, $L_h$, and $w_h$, a multiparemeter sweep was performed using COMSOL MP, and the results are shown in Fig.~\ref{fig:ATh}. In the left panel, the results include the findings from Fig.~\ref{fig:E_contours}, that is, for a fixed heater length (same line color) and same driving power, heaters with narrow width provide larger temperature gradients. Furthermore, the results for a fixed heater width (same symbols) show how, for a given driving power, increasing the length results into a redistribution of the driving power, and the local temperature increase is reduced. The latter, in terms of the accumulated phase shift (that is, $\Delta T_c \cdot L_h$) is shown in the right panel within the same Fig.~\ref{fig:ATh}. As expected, the configuration with more accumulated temperature is that with the narrowest width, $w_h=5$~\um, and longest length, $L_h=270$~\um. However, this not surprising remark is not the most important message from this graph. Instead, the trade-off between local temperature increase vs. accumulated increase vs. footprint  is the most noteworthy. Note this also in  Fig.~\ref{fig:ATh}(b), how in terms of $\Delta T_c \cdot L_h$, the combinations of length width 270x8~\ums and 170x5~\ums are equivalent. But this is brought in relation to Fig.~\ref{fig:B_rho}, since footprint reduction comes at the cost of higher increases resistance vs. power, and hence higher risk of damaging (burning) the thermal tuner.
 
As mentioned in the introduction, one figure typically employed to asses on the optical performance of the thermal tuner is the electrical power required for an optical phase change $\Delta \phi$ of $\pi$, named $P_\pi$. In order to derive the $P_\pi$ value from measurements, the optical spectral response for the MZIs having different tuner widths and lengths were recorded, for different input currents. The chip was held in a vacuum copper chuck, thermally stabilized to 25~\Celsius. Light from a C-band broadband source NP~Photonics~ASE-CL-20-S was injected in the MZI, and the optical spectrum was recorded with an OSA, Yokogawa~AQ6370C, with 10~pm resolution and sensitivity mode high. The asymmetric MZIs spectral response, with a path length difference of 622.38~\um, have a Free Spectral Range (FSR) of 2~nm, for the waveguide cross-section group index $n_g=1.93$. The spectral shifts for each drive current at the tuner were recorded, and from them, the corresponding phase shift derived, since for the MZI eq. (\ref{eq:phi_pi}) $\Delta \phi$ is related to the wavelength shift over the FSR as:
\begin{equation}
\label{eq:dwvl_fsr}
\frac{\Delta \phi}{2\pi} = \frac{\Delta\lambda}{\Delta\lambda_{FSR}}
\end{equation}
\begin{figure}
  {\par\centering
    \subfigure[$P_\pi$ for $L_h=270$~\um]{\resizebox*{0.48\textwidth}{!}{\includegraphics*{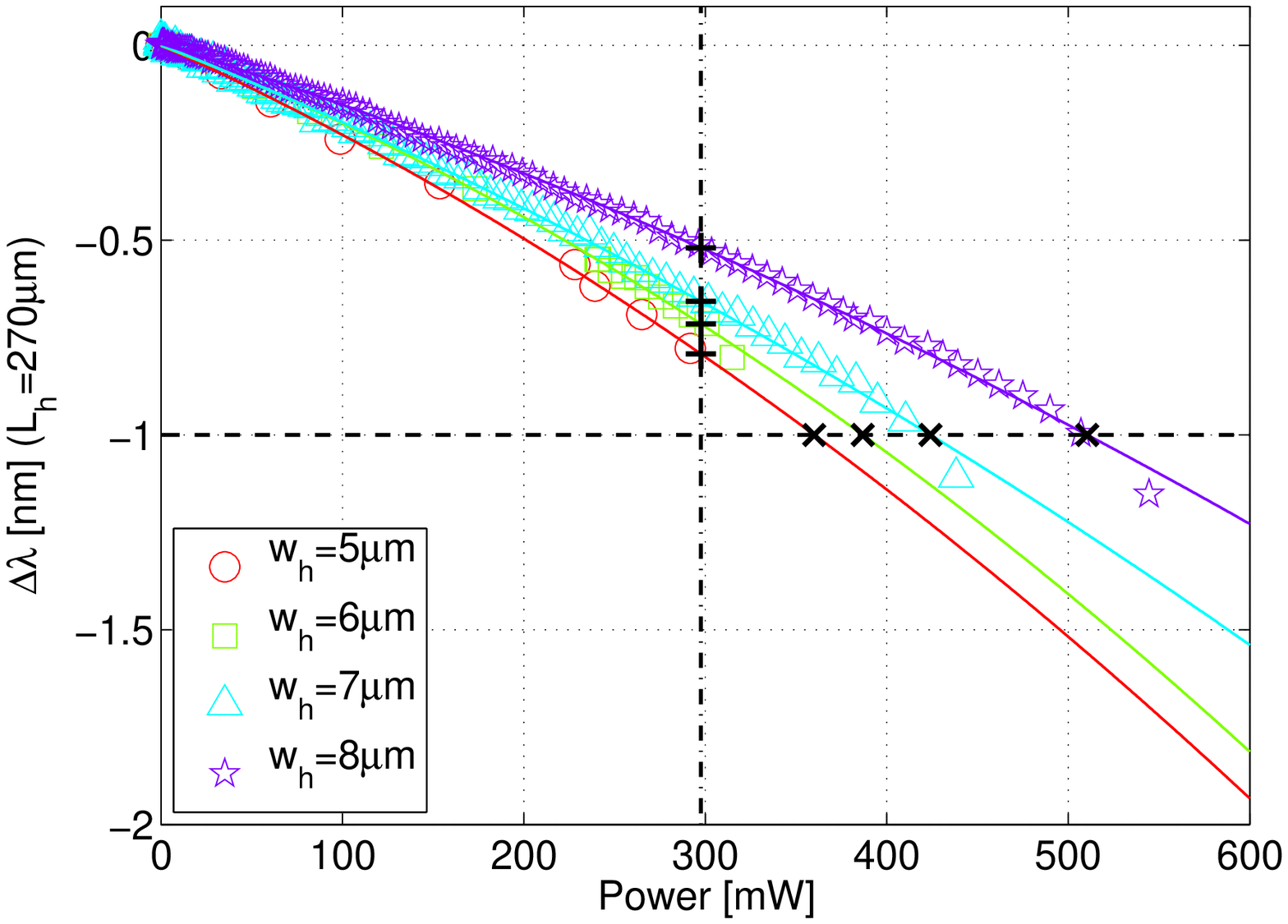}}}
    \subfigure[$P_\pi$ vs $L_h$]{\resizebox*{0.48\textwidth}{!}{\includegraphics*{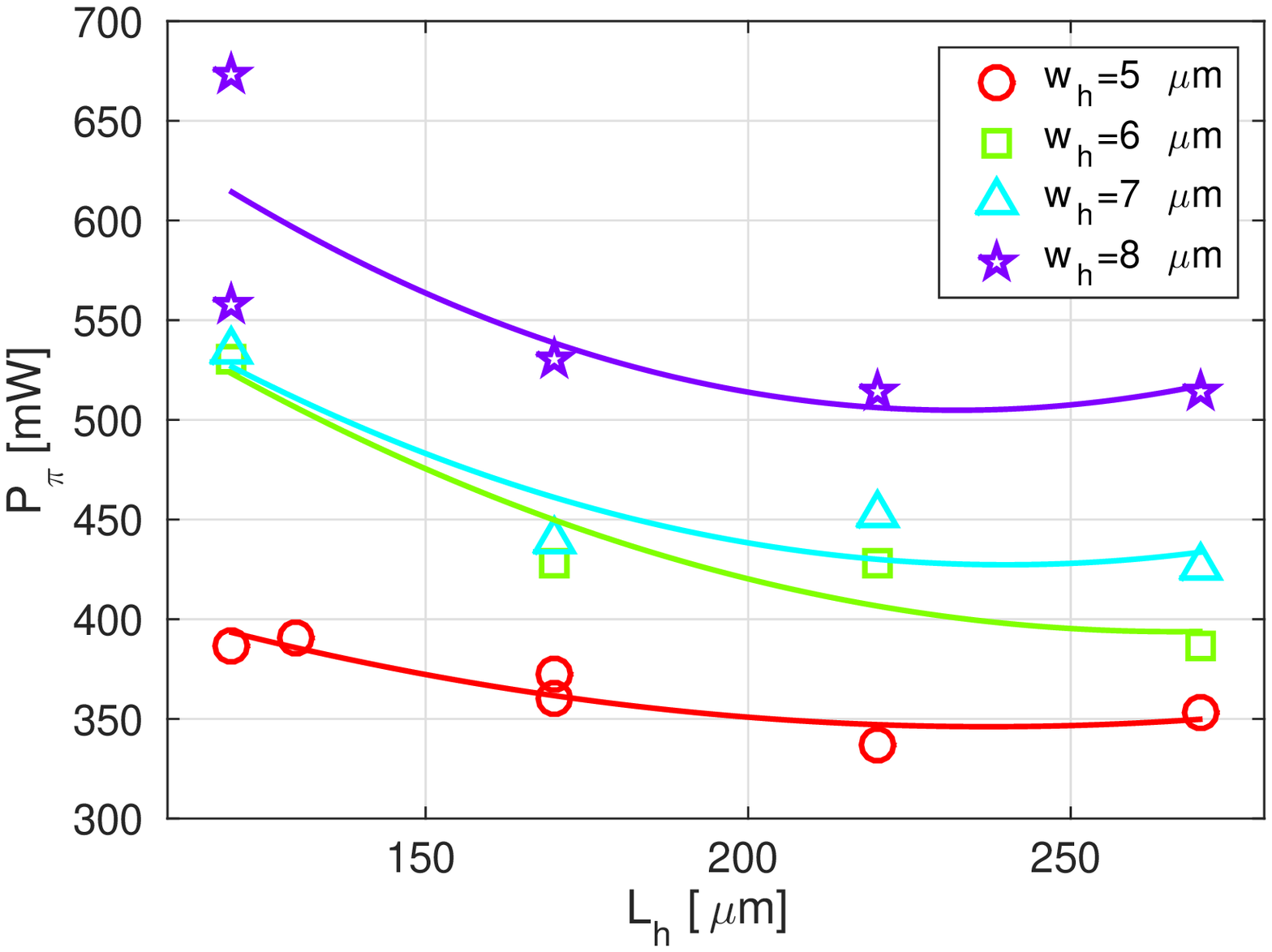}}}
    \caption{Estimated power required for a $\pi$ phase shift, vs thermal heater length, for different heater widths. Note that heaters that actually survived the tests are those of lengths above 270~\um, for every width.}
  \label{fig:D_Ppi}
  }
\end{figure}
All the shifts below the OSA resolution were discarded. The results are shown in Fig.~\ref{fig:D_Ppi}. Panel~(a) shows the wavelength shift for different drive powers, using heaters of length 270~\um~and different widths. As previously illustrated in Fig.~\ref{fig:E_contours} reducing the width increases the temperature gradient, therefore less power is required to generate the amount of heat for the $\pi$ phase shift (horizontal dashed line, cross symbols). Analog, for a given drive power, the shift is always larger for narrow heaters, as marked with a vertical dash-dot line and plus signs. In particular, this vertical line is set to the power of 297.62~mW of Fig.~\ref{fig:E_contours}. The wavelength shifts for widths from 5, 6, 7 and 8~\um, are -0.7922, -0.7153, -0.6574, -0.5197~nm, respectively. Using the equations (\ref{eq:phi_pi})-(\ref{eq:dwvl_fsr}), the temperature changes in the core $\Delta T_c(P=297.62mW,L_h=270\mu m,w_h)$ from Fig.~\ref{fig:E_contours}, and  $\partial \bar{n}/\partial T_c\simeq 3.05 \cdot 10^{-5}$~K$^{-1}$, the effective length over which the thermal effect takes place can be calculated with eq. (\ref{eq:phi_pi}). For this particular case, $L_h^e/L_h$ is 1.1834, 1.1220, 1.0667 and 0.9058 for widths from 5 to 8~\um.

Finally, the results in Fig.~\ref{fig:D_Ppi}(b) show how the estimated required $P_\pi$ increases progressively towards shorter heater lengths (smaller footprint). This is related to the effect of heat increasing the resistivity already discussed. Note that devices with lengths below 270~\um were extremely unstable and tended to get burnt easily, therefore concluding a length of 270~\um is safe in terms of heater resilience.

\section{\label{sec:xt}Thermal cross-talk}
\begin{figure}
  {\par\centering
   \resizebox*{0.48\textwidth}{!}{\includegraphics*{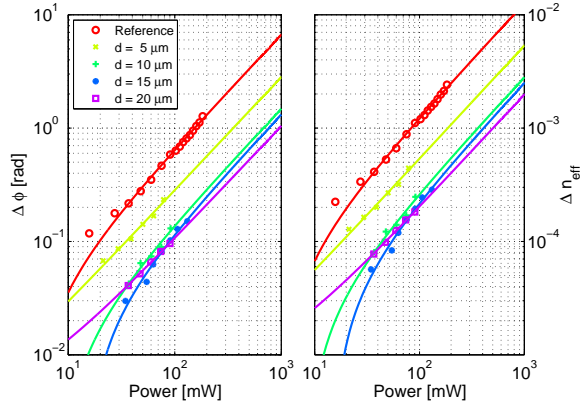}}
   \caption{Phase shift [rad] and equivalent effective index change ($\lambda$=1.55~\um) for the reference MZI (red circles) and for neighbor MZIs located at distances 5, 10, 15 and 20~\um~from the heater ($L_h=130$~\um, $w_h=5$~\um) -see Fig.~\ref{fig:chip}. Note the sign for shifts and changes in the reference MZI is opposite to the other, but is omitted to show log scale graphs.}
  \label{fig:Xt_near}
  }
\end{figure}

The heat generated by the thermal tuner of a MZI, propagates in the lateral direction and reaches other photonic devices on the chip, as can be also appreciated in Fig.~\ref{fig:E_contours}. For long thermal tuners with wider tracks, i.e. small $\Delta T_c$, no issues have been reported in the literature. The thermal cross-talk for nearby MZIs, Fig.~\ref{fig:chip} was investigated as follows: the thermal tuner of one MZI was biased and taken as reference, and the spectral traces recorded for both the biased MZI and the adjacent one. Once more from the spectral traces, the wavelength shifts were extracted. Then, owing to eq. (\ref{eq:dwvl_fsr}) the wavelength shifts were translated into their corresponding phase shifts and effective index change. The results are shown in Fig.~\ref{fig:Xt_near}, log scale, along with linear fitting lines. They correspond to the pairs of MZIs in Fig.~\ref{fig:chip} labeled as 5, 10, 15 and 20. All these MZIs have heaters with $w_h=5$~\um~and $L_h=130$~\um (compact and providing high $\Delta T_c$. From the results, two conclusions are derived. Firstly and as expected, increasing the distance between reduces the amount of cross-talk. The distance required for an influence of less than one order of magnitude is above 20~\um. Secondly, the relative distance between curves is constant with increasing powers, leading to the conclusion that the heat conduction through the materials is independent of the drive power.
\begin{figure}
  {\par\centering
    \subfigure[$L_h=220$~\um, $w_h=8$~\um, $d_1=86$~\um.]{\resizebox*{0.5\textwidth}{!}{\includegraphics*{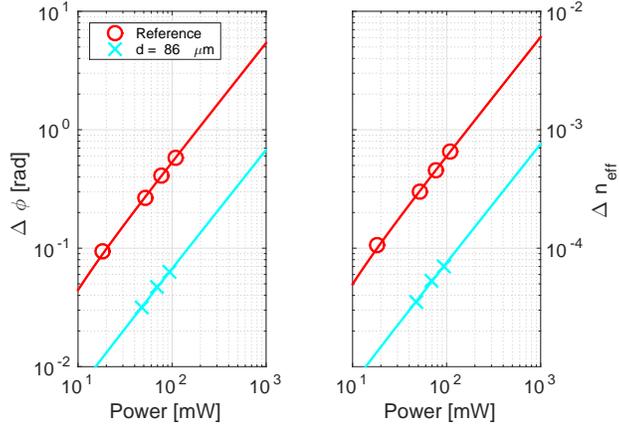}}}
    \subfigure[$L_h=120$~\um, $w_h=7$~\um, $d_2=426$~\um.]{\resizebox*{0.5\textwidth}{!}{\includegraphics*{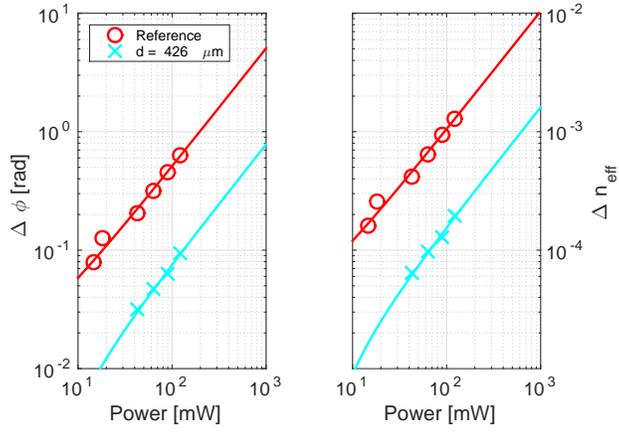}}}
    \subfigure[$L_h=270$~\um, $w_h=7$~\um, $d_3=586$~\um.]{\resizebox*{0.5\textwidth}{!}{\includegraphics*{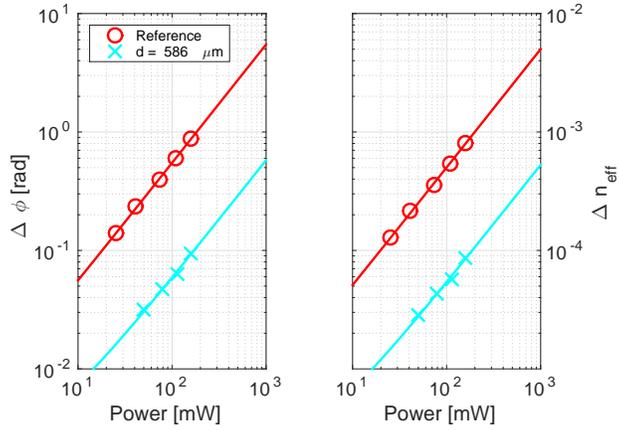}}}
   \caption{Phase shift [rad] and equivalent effective index change ($\lambda$=1.55~\um) for reference MZI (red circles) and for neighbor MZI arms located at several distances. The different panels correspond to thermal tuners with different lengths and widths}
  \label{fig:xt_other}
  }
\end{figure}

Albeit the chip layout was originally devised for the nearby cross-talk measurements shown, a few other structures allowed to perform similar measurements, for longer distances, labeled as $d_1$, $d_2$ and $d_3$ in Fig.~\ref{fig:chip}. The heaters in these structures have different widths and lengths of those in Fig.~\ref{fig:Xt_near}. Furthermore, as can be seen in Fig.~\ref{fig:chip}, there is not only the Silicon Oxide cladding between the tuned and affected MZIs, but also some other Silicon Nitride waveguides run through. All this prevents from being able to directly compare (i.e. include in the same graph) all these results among them. This is also supported by the findings in section~\ref{sec:ppi}, where the dependence of the temperature gradient (and hence the associated effective index and phase change) with heater structural (width and length) and operational (power) parameters were reported. Despite all the above, the results are provided in standalone individual graphs per heater configuration in Fig.~\ref{fig:xt_other} as reference.

\begin{figure}
  {\par\centering
    \resizebox*{0.52\textwidth}{!}{\includegraphics*{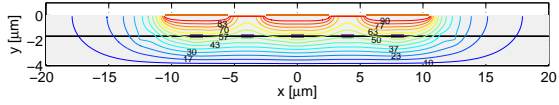}}
   \caption{Waveguide cross-section with multi-heater design on top, and array of waveguides with a pitch of 4~\um, with temperature gradient distribution overlaid. A dotted line is drawn from left to right crossing the core of the waveguides at half their height.}
  \label{fig:mejora}
  }
\end{figure}

\section{\label{sec:mejoras}Discussion and outlook}
In terms of the heat generated by the thermal tuner, all except the local temperature gradient surrounding the waveguide core is lost. Therefore, it may be of interest to explore novel heater-waveguide configurations, which do not resort to additional process steps. Just as a proposal, Fig.~\ref{fig:mejora} shows a cross-section with three metal stripes, and five cores beneath, corresponding to the same optical waveguide, which from top view perspective may be folded in different shapes such as spiral, serpentine or other. The dimensions and locations of the heaters and waveguides are engineered to fulfill two requirements. Firstly, the waveguides need to be placed apart a distance for which, when running in parallel, the optical cross-talk is negligible. In the picture, the waveguides are located at a distance of 4~\um, which according to the technology employed results into negligible optical cross-talk for as much as 40~mm parallel running length. Secondly, for all the optical waveguides to be subject to exactly the same temperature gradient, their position, along with the heater width and spacing, needs to be engineered as well. The picture shows a line across the waveguide cores, wherein the temperature gradient is the same. Furthermore note how three heaters can be used in this configuration to tune over five sections of the same waveguide. Therefore this provides an increase in efficiency of five to three, with respect to the case of a single heater-waveguide configuration. The multi-heater-waveguide array may be of use in layouts where long tunable delays are required, and for which energy consumption is an issue \cite{tech-sinx-mwp}. All this results into a more efficient thermal tuner, since a) waveguide section in between heaters benefit from otherwise lost heat b) the heater is folded so is also longer, therefore requiring less power consumption for a targeted index change and c) the overall heater structure uses less temperature gradient, so the cross-talk towards non-targeted waveguides will be smaller.


\section{\label{sec:conclusion}Conclusions}
The design trade-offs for the implementation of small footprint thermal tuners on silicon nitride have been presented. Test dies with Mach-Zehnder Interferometer test structure including compact thermal tuners of different dimensions were fabricated and tested, and the results were support with thermo-optic simulations. Compactness and thermal efficiency are traded off to heater deterioration and cross-talk. As path for improvement and future work, a multi-heater folded optical waveguide, where the number of heater stripes is less than the optical core cross-sections beneath, is proposed. Tuners with footprint of 270x5~\ums and switching power of 350~mW are reported among other, where the thermal-cross talk on dummy MZIs at distances over 20~\um, in terms of induced phase change, is less than one order of magnitude of that of the tuned device.

\appendix
\section{\label{app:res}Resistance for a linearly tapered conductor}
The resistance for a linearly tapered conductor  is given by:
\begin{eqnarray}
R_l & = & R_S \int_0^L \frac{1}{w_0 + \left( w_0-w_L \right) \frac{z}{L}} \partial z \nonumber \\
& =  & R_S \frac{L}{w_L-w_0} \left( \log(w_L) - \log(w_0) \right)
\end{eqnarray}
Hence, $F_l$ and $F_a$ are given by:
\begin{equation}
\label{eq:Fl}
F_l = \frac{L_l}{W_l-w_a}\left(\log(W_l)-\log(w_a)\right)
\end{equation}
\begin{equation}
\label{eq:Fa}
F_a = \frac{\pi r_a / 2 }{w_a-w_h}\left(\log(w_a)-\log(w_h)\right)
\end{equation}

\section{\label{app:mat}Summary of material properties}
For the materials involved, the following parameters were used. Refractive indices:
\begin{equation}
n_{SiO_2}\left(\lambda\left[\mu m\right]\right) = \sqrt{\frac{1.09877\lambda^2}{\lambda^2-92.4317^2}+1}
\end{equation}
\begin{equation}
n_{Si_3N_4} \left(\lambda\left[\mu m\right]\right) = \sqrt{\frac{2.8936\lambda^2}{\lambda^2+139.67^2}+1}
\end{equation}
Thermo-optic coefficients:
\begin{equation}
\frac{\partial n_{SiO_2}}{\partial T} = 8.6\cdot 10^{-6} \mathrm{K}^{-1}
\end{equation}
\begin{equation}
\frac{\partial n_{Si_3N_4}}{\partial T} = 4.5\cdot 10^{-5} \mathrm{K}^{-1}
\end{equation}
Thermal conductivity:
\begin{equation}
\sigma_{SiO_2} = 1.5 \frac{W}{m \cdot K}
\end{equation}
\begin{equation}
\sigma_{Si_3N_4} = 30.5 \frac{W}{m \cdot K}
\end{equation}
In COMSOL Multi-Physics, the measured resistivity reported in Section~\ref{sec:elec} was employed for the simulations.

\section*{Acknowledgment}
The authors acknowledge financial support by: the Spanish CDTI NEOTEC start-up programme, the Spanish MINECO projects TEC2013-42332-P PIF4ESP, TEC2015-69787-REDT PIC4TB, and TEC2014-60378-C2-1-R MEMES, the GVA PROMETEO 2013/012 research excellency award, the EU H2020-ICT-27-2015 CSA Proposal number 687777, project FEDER UPVOV 10-3E-492 and project FEDER UPVOV 08-3E-008. D. Pérez acknowledges final support through the FPI grant programme by Universitat Politècnica de València.


\bibliographystyle{IEEEtran}
\bibliography{dperez_sinx_thermo.bib}

\end{document}